\documentclass[aps,pra,twocolumn,showpacs,superscriptaddress]{revtex4-1}
\usepackage[colorlinks ,linkcolor=blue,anchorcolor=blue,citecolor=blue,urlcolor=blue]{hyperref}
\usepackage{graphicx}
\usepackage{epstopdf}
\usepackage{soul, color, xcolor}
\usepackage{bm}
\usepackage{amsmath,amssymb,amsfonts}
\usepackage{appendix}

\def\be{\begin{equation}} \def\ee{\end{equation}}
\def\bea{\begin{eqnarray}} \def\eea{\end{eqnarray}}

\begin{document}

\title{Staggered nonlinear spin generations in centrosymmetric altermagnets under electric current}
\author{Jie Zhang}
\affiliation{School of Physics and Technology, Nanjing Normal University, Nanjing 210023, China}
\affiliation{Center for Quantum Transport and Thermal Energy Science,
Nanjing Normal University, Nanjing 210023, China} 
\author{Ruijing Fang}
\affiliation{School of Physics and Technology, Nanjing Normal University, Nanjing 210023, China}
\affiliation{Center for Quantum Transport and Thermal Energy Science,
Nanjing Normal University, Nanjing 210023, China} 
\author{Zhichao Zhou}
\affiliation{School of Physics and Technology, Nanjing Normal University, Nanjing 210023, China} 
\author{Xiao Li} 
\email{lixiao@njnu.edu.cn}
\affiliation{School of Physics and Technology, Nanjing Normal University, Nanjing 210023, China}
\affiliation{Center for Quantum Transport and Thermal Energy Science,
Nanjing Normal University, Nanjing 210023, China}

\begin{abstract} 
Current-induced spin generations are of significant importance for electrically controllable magnetization. Due to symmetry constraints, linear spin generation is absent in centrosymmetric magnets and nonlinear contributions become crucial. However, nonlinear spin generations have few examples in centrosymmetric compensated magnets with opposite-spin sublattices, which hinders electric control of associated magnetization. Here, we study nonlinear spin generations in altermagnets with opposite-spin sublattices. In a square altermagnetic model, both staggered and uniform nonlinear spin generations appear at opposite-spin sublattices. They vary as the magnetization direction rotates, with emerging out-of-plane components that can be utilized in perpendicular magnetization switching of high-density storage devices. By first-principles calculations, out-of-plane, staggered nonlinear spin generations are found to be considerable in a typical altermagnet, Fe$_2$Se$_2$O monolayer. Our findings provide opportunities for electrically manipulating magnetization and designing energy-efficient magnetic devices based on compensated magnets.
\end{abstract} 
\maketitle

{\color{blue}\textit{Introduction.}} -- Electric control of magnetization has recently received significant attentions.
In magnets, current-induced spin generations and associated torques play a role of effective magnetic field, which has enormous applications in spintronics, e.g. driving magnetization switching and domain-wall motion \cite{RevModPhys.91.035004,Chappert_Fert_and_Dau_2007,Brataas_Kent_and_Ohno_2012}.
Specific to antiferromagnets, it has been theoretically recognized that staggered spin generations on antiferromagnetically coupled sites are essential to the manipulation of magnetization dynamics  \cite{Zelezny_2014,Zelezny_2017}, and the reorientation of antiferromagnetic Néel vector has then been realized under the action of electric currents in experiments \cite{Wadley_2016,Bodnar_Smejkal_2018,Zhou_Zhang_li_2018,Selzer_Salemi_Deak_2022}.
Previous studies have primarily focused on magnetic materials with inversion symmetry breaking that is regarded as a necessary condition for linear spin generations under electric current \cite{Zelezny_2014,Zelezny_2017}. 
When spin generation has been recently extended to nonlinear regime, electric control of magnetization is also expected in centrosymmetric magnets through nonlinear spin generations  \cite{Xiao_2022,Xiao_2023}.
However, compared with ferromagnets, there have been few studies on nonlinear spin generations in compensated magnets with opposite-spin sublattices, and site dependence of spin generations remains elusive.

On the other hand, altermagnetism, as a new type of compensated magnetic phase, exhibits exotic characteristics \cite{Hayami_2019,Yuan_2020,Hayami_2020,Liu_2022,Smejkal_2022,Smejkal_Sinova_2022}.
The altermagnets have opposite-spin sublattices that are connected by rotational or mirror symmetry in real space, while spin-splitted bands appear in momentum space.
For the new magnetic phase, the primary method to control magnetization reorientation still relies on applying magnetic fields \cite{Han_2024,Urata_2024}, with few studies on electric control. Therefore, centrosymmetric altermagnets with opposite-spin sublattices provide a fertile platform for studying nonlinear spin generations and associated site dependence, which are expected to be utilized in electric control of magnetization in altermagnets.

In this work, we study nonlinear spin generations in centrosymmetric altermagnets, by combining symmetry analyses, tight-binding model and first-principles calculations.
In a two-dimensional square altermagnetic model with orbital orderings, both staggered and uniform spin generations are found on two opposite-spin sublattices.
Their magnitudes and signs vary with the magnetization direction, with emerging out-of-plane staggered spin components.
The staggered spin generations can be used to regulate magnetic dynamics of opposite-spin sublattices, while the out-of-plane spin generations are particularly desirable for high-density storage devices based on out-of-plane anisotropic magnets.
Taking Fe$_2$Se$_2$O monolayer as a typical example of two-dimensional altermagnets, first-principles calculations of nonlinear spin generations are then performed, and corresponding results support the findings from the model.
Out-of-plane staggered nonlinear spin generations in the monolayer reach the order of  $10^4$ $\mu _B/$V$^{2}$, which is larger than that of previous studied ferromagnets.
Our studies extend nonlinear spin generations to compensated magnets, and pave the way for manipulating magnetization of opposite-spin sublattices by electric means.

\begin{figure}[htb]
\includegraphics[width=8.0 cm]{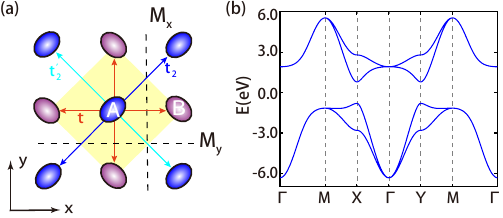}
\caption{ Orbital ordering and band structure of a altermagnetic square lattice. (a) Local atomic orbitals in  a square lattice. The blue and purple ellipses stand for two mutually perpendicular orbitals on $A$ and $B$ sublattices, which leads to anisotropic next-nearest-neighbor hoppings denoted by $t_2$ and $t_2^{'}$. The $A$ and $B$ sublattices are related by mirror symmetries, $\mathcal{M}_{x/y}$. The yellow shade represents a unit cell of the lattice.
(b) Band structure of the lattice. The parameters used here satisfy relations, $t=t_2=J_\text{ex}=1$ eV, $t_2^{'} = 0.1$ eV, and $t_\text{soc}=0.8$ eV.
}
\label{fig1}
\end{figure}

{\color{blue}\textit{Crystal structure and symmetry analyses.}} -- 
We first consider a two-dimensional altermagnetic model of a square lattice with anisotropic hopping parameters \cite{Brekke_2023,Gustavo_2024,Attias_2024}, which is shown in Fig. \ref{fig1}a. The square lattice includes two inequivalent sublattices, denoted as $A$ and $B$, which have different orbital orderings. That is, two anisotropic orbitals at $A$ and $B$ sublattices exhibit the same shape but with mutually perpendicular orientations. Moreover, considering magnetic ordering, there are opposite magnetic moments at $A$ and $B$ sublattices.
From the view of the crystal symmetry, the whole lattice, as well as $A$ and $B$ sulattices, has the inversion symmetry $\mathcal{P}$. Owing to the presence of the orbital ordering, the $A$ and $B$ sublattices can't be connected by the combined symmetry of $\mathcal{P}$ and time reversal  $\mathcal{T}$, which leads to the altermagnetic order \cite{Brekke_2023}.
Besides, there are two mirror planes, $\mathcal{M}_x$ and $\mathcal{M}_y$, of which the normal direction are along the $x$-axis and $y$-axis, respectively.  They connect the two sublattices.

Based on the above symmetries, symmetry analyses of current-induced spin generations, $\delta \textbf{\textit{s}}$, in the square lattice are performed. Firstly, the presence of the $\mathcal{P}$ symmetry excludes the linear spin response to applied electric field, i.e. $\delta \textbf{\textit{s}}\propto E$, where $E$ represents an applied electric field. It is because under the $\mathcal{P}$ operation, $\delta \textbf{\textit{s}}$ is invariant, but $E$ is reversed, rendering the linear relationship to no longer hold. While the linear spin response is absent, the quadratic spin response is likely to become the leading effect.
The quadratic spin response to applied electric fields can be expressed as
\begin{equation}\label{tab:Eq1}
\delta s_a= \chi  _{abc}E_bE_c,
\end{equation}
where $ \chi_{abc} $ is the nonlinear response tensor, with the subscripts, $a$, $b$, $c$, represents Cartesian coordinates. While $a$ denotes the spin direction, $b$ and $c$ denote directions of electric fields.
The Einstein summation convention is adopted in the equation. As for the quadratic spin response, the product of two electric-field components, as well as $\delta \textbf{\textit{s}}$, is invariant under the $\mathcal{P}$ operation. The response is thus symmetry-allowed in the centrosymmetric lattice. Further considering that the quadratic spin responses have two contributions, i.e. $\mathcal{T}$-odd part from the Berry curvature dipole \cite{Xiao_2022} and $\mathcal{T}$-even part from the anomalous spin polarizability dipole \cite{Xiao_2023}, $\chi_{abc}$ can be divided into $\chi_{abc}^{\text{o}}$ and $\chi_{abc}^{\text{e}}$. Besides, when distinguishing $A$ and $B$ sublattices, an extra subscript, $i=A$ or $B$, is also added into the tensor $\chi$ and the spin generation $\delta \textbf{\textit{s}}$. The tensor element $\chi _{abc,i}^{\text{o/e}}$ obeys the symmetry transformation rule \cite{Xiao_2022,Xiao_2023},
\begin{equation}
\begin{split}\label{tab:Eq2}
\chi_{a^{'}b^{'}c^{'},i^{'}}^{\text{o/e}}=\eta _\mathcal{T}\text{det}(O)O_{a^{'}a}O_{b^{'}b}O_{c^{'}c}\chi _{abc,i}^{\text{o/e}}.
\end{split}
\end{equation}
Here, $O$ is a unitary matrix of a point group operation, which relates the sublattices $i$ and $i^{'}$. When $\chi_{abc,i}^{\text{o}}$ is considered, the factor $\eta _\mathcal{T}=+  1$ for a nonprimed operation, and $\eta _\mathcal{T}=-  1$ for a primed operation that combines a spatial operation and $\mathcal{T}$.
As for $\chi_{abc,i}^{\text{e}}$, $\eta _\mathcal{T}=+  1$ for both nonprimed and primed operations.

\begin{table}
\caption{\label{tab:table1} Symmetry constrains on tensor elements of nonlinear spin responses. The first and second lines list directions of local magnetic moments and corresponding magnetic point group symmetries, respectively. For each tensor element from the third to the eighth line, "$+$","$-$" and "$\times $" denote uniform, staggered and vanishing spin generations on two sublattices, respectively.
}
\begin{ruledtabular}
\begin{tabular}{cccc}
$ $&$M\parallel x$&$M\parallel y$ &$M\parallel z$\\
\hline
$\hat{O}$& $\mathcal{M}_y,C_2^z\mathcal{T}$&$\mathcal{M}_x,C_2^z\mathcal{T}$&$\mathcal{M}_x,\mathcal{M}_y,C_2^z\mathcal{T}$\\
 $\chi_{xxx}^{\text{o/e}}$& $-/\times $& $+/\times $ &$\times/\times $\\
$\chi_{yxx}^{\text{o/e}}$&$+/\times $& $-/\times $ &$\times/\times $\\
$\chi_{zxx}^{\text{o/e}}$&$\times/- $& $\times/-$ &$-/- $\\
$\chi_{xyy}^{\text{o/e}}$& $-/\times $&$+/\times $ &$\times/\times $\\
$\chi_{yyy}^{\text{o/e}}$& $+/\times $& $-/\times $ &$\times/\times $\\
$\chi_{zyy}^{\text{o/e}}$& $\times/- $& $\times/-$ &$-/- $\\
\end{tabular}
\end{ruledtabular}
\end{table}

According to Eq \ref{tab:Eq2}, possible spin generations on $A$ and $B$ sublattices are analyzed for highly-symmetric directions of local magnetic moments, as listed in Table \ref{tab:table1}. In the table, the symmetry operations connecting different sublattices, $\mathcal{M}_{x/y}$, are used to explore the relationship between $\delta s_A$ and $\delta s_B$, while the symmetry operation, $C_2^z\mathcal{T}$ or $C_2^z$,  transforms the sublattices into themselves and determines the presence or absence of the spin generations. Further considering the setup of two-dimensional transport experiment, applied electric fields are restricted to the $x-y$ plane and along the same direction. When the local magnetic moments are aligned along the $x$ axis, it is found that there are staggered spin generations on two sublattices, i.e. $\delta s_A=-\delta s_B$, for $\chi_{xxx}^{\text{o}}$, $\chi_{xyy}^\text{o}$, $\chi_{zxx}^{\text{e}}$, and $\chi_{zyy}^\text{e}$. 
Besides, spin generations on two sublattices are uniform, i.e. $\delta s_A=\delta s_B$, for $\chi_{yxx}^{\text{o}}$, and $\chi_{yyy}^\text{o}$. 
As for the local magnetic moments along the $y$ axis, staggered and uniform spin generations are also found for different tensor elements of nonlinear spin response. On the other hand, when the local magnetic moments are along the $z$ axis, there are only staggered, out-of-plane spin generations on two sublattices for both electric fields along $x$ and $y$ axes.

According to the above symmetry analyses, there are two points worthy of attentions. Firstly, staggered nonlinear spin generations are symmetry-allowed, and they are highly desirable for effective switching of local magnetic moments in centrosymmetric compensated magnets. This is analogous to the role of the linear spin generation in noncentrosymmetric antiferromagnets \cite{Shao_2023,Zhang_2022}. Besides, uniform spin generations found here are likely to be used as a spin source to reorient magnetic moments of adjacent materials. 
Secondly, out-of-plane spin components can be generated, which is crucial for field-free  perpendicular magnetization reversal of out-of-plane anisotropic magnets in high-density storage devices \cite{Shuai_2022,You_2021,Engel_2007}.

{\color{blue}\textit{Model calculations of  nonlinear spin generations.}} -- 
Armed with the symmetry analyses, we then employ a tight-binding Hamiltonian of the altermagnetic square lattice to calculate electronic band structure and spin generations. Using $\{\psi_{A,\uparrow }, \psi_{A,\downarrow }, \psi_{B,\uparrow }, \psi_{B,\downarrow }\}$ as a basis set, the Hamiltonian reads, 
\begin{equation}
\begin{aligned}\label{tab:Eq3}
H=&t\displaystyle\sum_{<i,j>,\alpha }c_{i,\alpha}^{\dagger }c_{j,\alpha }+t_2\displaystyle\sum_{\ll i,j\gg,\alpha}c_{i,\alpha}^{\dagger }c_{j,\alpha }\\
+&t_2^{'}\displaystyle\sum_{\ll i,j\gg,\alpha}c_{i,\alpha}^{\dagger }c_{j,\alpha }-J_\text{ex}\displaystyle\sum_{i,\alpha ,\beta}\bm{S}_ic_{i,\alpha}^{\dagger }\bm{\sigma}_{\alpha \beta}c_{i,\beta}\\
+&it_\text{soc}\displaystyle\sum_{<i,j>,\alpha,\beta}\nu _{ij}c_{i,\alpha}^{\dagger }\sigma^z_{\alpha \beta}c_{j,\beta}.
\end{aligned}
\end{equation}
In the basis set, $\psi_{i,\alpha }$ denotes the wave function of the sublattice $i$ with spin $\alpha$ $(\alpha=\ \uparrow$ or $\downarrow)$. The Hamiltonian includes five terms, corresponding to nearest-neighbor (NN) hoppings, two kinds of next-nearest-neighbor (NNN) hoppings exhibiting anisotropy, exchange interaction between itinerant spins and local magnetic moments, and intrinsic spin-orbit coupling (SOC) from NN hoppings, respectively.
$c_{i,\alpha}^{\dagger} (c_{i,\alpha})$ is the creation (annihilation) operator of an electron. While $t$ represents the NN hopping strength, $t_2$ and $t_2^{'}$ denote strengths of anisotropic NNN hoppings, as illustrated in Fig. \ref{fig1}a. $J_\text{ex}$ and $t_\text{soc}$ quantifies strengths of the exchange interaction and SOC, respectively. $\bm{S}_i$ represents the unit vector of local magnetic moment, with $\bm{S}_A=-\bm{S}_B$ for the altermagnet with opposite-spin sublattices. $\bm{\sigma}=(\sigma_x,\sigma_y,\sigma_z)$ are spin Pauli matrices. For the horizontal NN hoppings from $A(B)$ sublattice to $B(A)$ sublattice in Fig. \ref{fig1}a, $\nu _{ij}$ in the SOC term adopts $+1(-1)$. As for the vertical hoppings, $\nu _{ij}$ reverses its sign.


Fig. \ref{fig1}b demonstrates the band structure of the altermagnetic square lattice, with the magnetization along the $z$ axis. It is seen that there are four bands, indicating the spin degeneracy is lifted for the altermagnetic order. Besides, for the parameters considered here, the valence band maximum and conduction band minimum are located at the high-symmetric $X$ and $Y$ points, forming two valleys for both bands. The two valleys are degenerate, which is protected by the mirror symmetries, $\mathcal{M}_{x/y}$ \cite{Ma_2021}.
The tight-binding band structure is similar to the first-principle one of the  Fe$_2$Se$_2$O monolayer, which will be studied later \cite{Wu_2024}.


\begin{figure}[htb]
\includegraphics[width=8.5 cm]{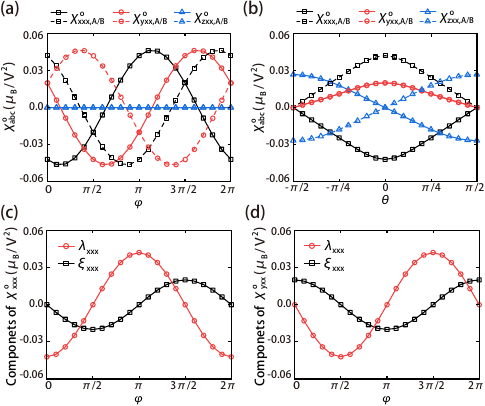}
\caption{Nonlinear spin generations in a altermagnetic square lattice. 
(a) Evolutions of $\chi^{\text{o}}_{xxx,i}$, $\chi^{\text{o}}_{yxx,i}$ and $\chi^{\text{o}}_{zxx,i}$ as functions of  the azimuthal angle $\phi$ with fixed $\theta =0$. 
(b) Evolutions of $\chi^{\text{o}}_{xxx,i}$, $\chi^{\text{o}}_{yxx,i}$ and $\chi^{\text{o}}_{zxx,i}$ as functions of  the canting angle $\theta$ with fixed $\phi =0$. 
(c) and (d) The staggered contribution, $\lambda_{abc}$, and the uniform contribution, $\xi_{abc}$, as functions of $\phi$ for $\chi_{xxx}^{\text{o}}$ and $\chi_{yxx}^{\text{o}}$, respectively. The chemical potential is set to 1 eV.
}
\label{fig02}
\end{figure}

According to the tight-binding Hamiltonian \ref{tab:Eq3}, we further calculate quadratic spin generations, as shown in Fig. \ref{fig02}, while linear spin generations are verified to be zero. The calculation details are provided in the Supporting Information (SI hereafter).
In Fig. \ref{fig02}, $\mathcal{T}$-odd part of the nonlinear spin generations, $\chi_{abc,\gamma}^{\text{o}}$, are presented as functions of the direction of local magnetic moment, with the electric field along the $x$ axis, i.e. $b=c=x$ in Eq. \ref{tab:Eq1}, and with a fixed chemical potential, $\mu =\  $1 eV.
The direction of the local magnetic moment at the $A$ sublattice, $\bm{S}_A$, is described by two angles, i.e. the azimuthal angle, $\phi$, and the canting angle, $\theta$. Given $\bm{S}_A=-\bm{S}_B$ for the altermagnet, $\phi$ and $\theta$ denote the direction of the Néel vector as well.
Figs. \ref{fig02}a and \ref{fig02}b demonstrate evolutions of the $\mathcal{T}$-odd part with varied $\phi$ in the equatorial plane ($\theta =0$) and with varied $\theta$ in the meridional plane passing through the $x$ axis ($\phi =0$), respectively. 

In Fig. \ref{fig02}a, it is seen that the in-plane spin generations, $\chi_{xxx,A}^{\text{o}}$, $\chi_{xxx,B}^{\text{o}}$, $\chi_{yxx,A}^{\text{o}}$, and $\chi_{yxx,B}^{\text{o}}$, exhibit trigonometric dependence on $\phi$ with nonzero initial phases, while the out-of-plane spin generations, $\chi_{zxx,A}^{\text{o}}$ and $\chi_{zxx,B}^{\text{o}}$ are vanishing. When $\phi = 0$, $\pi$ or $2\pi$, $\chi_{xxx,A}^{\text{o}}=-\chi_{xxx,B}^{\text{o}}$, corresponding to staggered spin generations, and  $\chi_{yxx,A}^{\text{o}}=\chi_{yxx,B}^{\text{o}}$, corresponding to uniform spin generations. When $\phi = \pi/2$ or $3\pi/2$, $\chi_{xxx}^{\text{o}}$ and $\chi_{yxx}^{\text{o}}$ become uniform and staggered ones, respectively.
The results agree with the symmetry analyses in Table \ref{tab:table1}.
Besides, when $\phi$ takes on other values, in-plane spin generations on $A$ and $B$ sublattices are neither equal nor opposite, indicating that there are both staggered and uniform contributions. We thus divide the spin generations into staggered contribution, $\lambda_{abc}$, and uniform contribution, $\xi_{abc}$,
i.e. $\chi_{abc,A}^{\text{o}}=\lambda _{abc}+\xi_{abc}$, $\chi_{abc,B}^{\text{o}}=\xi_{abc}-\lambda_{abc}$.
Figs. \ref{fig02}c and \ref{fig02}d present $\lambda_{abc}$ and $\xi_{abc}$ as functions of $\phi$ for $\chi_{xxx}^{\text{o}}$ and $\chi_{yxx}^{\text{o}}$, respectively. For $\chi_{xxx}^{\text{o}}$ ($\chi_{yxx}^{\text{o}}$), the staggered and uniform contributions look like cosine (sine) and sine (cosine) functions of $\phi$ without any phase shift.
The amplitude of the staggered contribution is about twice than that of the uniform one. Therefore, both the staggered and uniform spin generations act on two sublattices of the altermagnet, and they jointly determine magnetization dynamics of the altermagnet. 

In Fig. \ref{fig02}b, with varied canting angle $\theta$ in the meridional plane passing through the $x$ axis, $\chi_{xxx}^{\text{o}}$, $\chi_{yxx}^{\text{o}}$ and $\chi_{zxx}^{\text{o}}$ are all nonvanishing.
$\chi_{xxx}^{\text{o}}$ and $\chi_{zxx}^{\text{o}}$ are always staggered on $A$ and $B$ sublattices, while $\chi_{yxx}^{\text{o}}$ is uniform, which are ensured by the mirror symmetry, $\mathcal{M}_y$, and can be derived by Eq. \ref{tab:Eq2}. When focusing on the staggered out-of-plane spin generations, $\chi_{zxx}^{\text{o}}$, their magnitudes are comparable to in-plane ones. Besides, the spin generations are opposite to local magnetic moments for the parameters used here. Therefore, they are likely to be utilized for perpendicular magnetization switching in high-density storage devices based on out-of-plane anisotropic magnets.

 In addition, we also calculate the $\mathcal{T}$-even part, $\chi_{abc,\gamma}^{\text{e}}$, of the quadratic spin generations, which is several orders of magnitude smaller than the $\mathcal{T}$-odd part. The case of the electric field along the $y$ axis is also considered and it is similar to that of the electric field along the $x$ axis. The results of the $\mathcal{T}$-even part and the electric field along the $y$ axis are given in SI.

\begin{figure}[htb]
\includegraphics[width=8.5 cm]{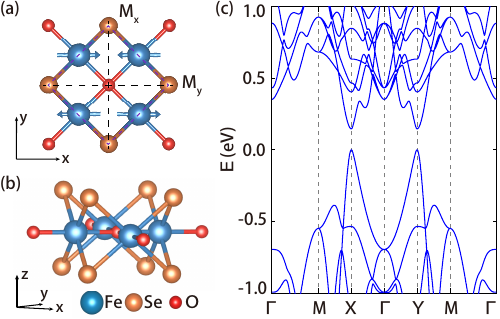}
\caption{Crystal and band structures of the Fe$_2$Se$_2$O monolayer.
(a) Top view and (b) side view of the crystal structure. Blue, yellow and red balls stand for Fe, Se and O atoms, respectively. A unit cell of the monolayer is bounded by purple dashed lines. The arrows represent opposite spin directions on two inequivalent Fe atoms in a unit cell.
(c) Band structure of the monolayer, with considering the spin-orbit coupling and the magnetization along the $z$ axis. The valence band maximum is set to zero energy. 
}
\label{fig03}
\end{figure}

{\color{blue}\textit{First-principles calculations of nonlinear spin generations} --}
Based on the above symmetry analyses and model calculations, nonlinear spins can be expected to be generated in a centrosymmetric altermagnet under electric current. By first-principles based calculations, we study nonlinear spin generations in a typical two-dimensional altermagnetic material, Fe$_2$Se$_2$O monolayer \cite{Wu_2024}.

Figs. \ref{fig03}a and \ref{fig03}b demonstrate top and side views of the crystal structure of the Fe$_2$Se$_2$O monolayer, respectively.
The monolayer has a tetragonal structure, with three atomic layers. While the middle layer contains Fe and O atoms, the atoms in both upper and lower layers are Se atoms and they are arranged symmetrically with respect to the middle layer. A unit cell of the monolayer includes one chemical formula unit, i.e. two Fe atoms, two Se atoms, and one O atom. Each Fe atom is coordinated by two O and four Se atoms. These neighboring atoms form a distorted octahedron, with calculated Fe-O and Fe-Se bond lengths being 2.02 \AA \
and 2.64 \AA, respectively.
The O-Fe-O bonds thus serve as the short axis of the distorted octahedron. It is seen that the directions of the short axes of the distorted octahedra surrounding two Fe atoms in a unit cell are perpendicular to each other, which is expected to create perpendicular orbital orderings at the two Fe atoms like the case of the square model in Fig. \ref{fig1}. Besides, these two Fe atoms are related by mirror symmetries $\mathcal{M}_{x/y}$, located on Se-O-Se planes. By the calculations of magnetic properties, they possess opposite magnetic moments with a magnitude of $\sim$  3.83 $\mu_B$. Therefore, the magnetic Fe atoms form an altermagnetic order.

Fig. \ref{fig03}c shows the first-principles band structure of the Fe$_2$Se$_2$O monolayer, with considering the SOC and the magnetization along the $z$ axis.
In the band structure, two degenerate valleys appear at $X$ and $Y$ points for both valence and conduction bands, similar to our tight-binding result. The size of direct band gaps at $X$ and $Y$ valleys is computed to be 146.5 meV.
The band structure is consistent with previous calculations \cite{Wu_2024}.

\begin{figure}[htb]
\includegraphics[width=8.5 cm]{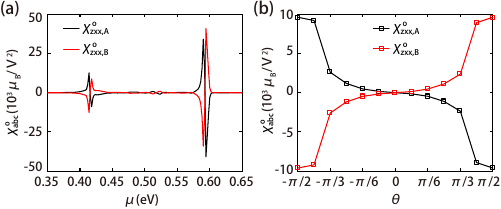}
\caption{Nonlinear spin generations in the Fe$_2$Se$_2$O monolayer.
(b) Evolutions of $\chi^{\text{o}}_{zxx,A}$ and $\chi^{\text{o}}_{zxx,B}$ as functions of  the chemical potential $\mu$, with fixed $\theta = \pi/2$ and $\phi =0$.
(b) Evolutions of $\chi^{\text{o}}_{zxx,A}$ and $\chi^{\text{o}}_{zxx,B}$ as functions of  the canting angle $\theta$, with fixed $\phi =0$ and $\mu=$0.42 eV.
The conduction band minimum is set to zero energy. 
}
\label{fig04}
\end{figure}

Nonlinear spin generations are then calculated by Wannier-based tight-binding models with first-principles inputs.
In the followings, we take $\mathcal{T}$-odd out-of-plane spin generations under the electric field along $x$ axis, i.e. $\chi^{\text{o}}_{zxx,A}$ and $\chi^{\text{o}}_{zxx,B}$, as examples, where $A$ and $B$ sublattices correspond to two inequivalent Fe atoms with opposite spins.
As mentioned above, the out-of-plane spin generations are closely related to perpendicular magnetization switching in high-density storage devices.
Fig. \ref{fig04}a presents $\chi^{\text{o}}_{zxx,A}$ and $\chi^{\text{o}}_{zxx,B}$ as functions of the chemical potential, $\mu$.
It is seen that $\chi^{\text{o}}_{zxx,A}$ and $\chi^{\text{o}}_{zxx,B}$ have the same magnitude but opposite signs, indicating staggered nonlinear spin generations and agreeing with the model calculations.
The magnitude of $\chi^{\text{o}}_{zxx,A}$ and $\chi^{\text{o}}_{zxx,B}$ varies with $\mu$. 
When $\mu$ is around 0.4 eV with respect to the conduction band minimum, $\chi^{\text{o}}_{zxx,A(B)}$ has two opposite neighboring peaks, with magnitudes of 1.3$\times 10^4$ $\mu_B/$V$^2$ and 0.9$\times 10^4$ $\mu_B/$V$^2$, respectively. 
These values are two orders of magnitude larger than nonlinear spin generations of ferromagnetic MnBi$_2$Te$_4$ single-layer \cite{Xiao_2022}.
Therefore, considerable spin generations are expected in the Fe$_2$Se$_2$O monolayer by electron doping.
Besides, further moving away from the conduction band minimum, higher peaks are found around 0.6 eV, with magnitudes of 3.2$\times 10^4$ $\mu_B/$V$^2$ and 4.0$\times 10^4$ $\mu_B/$V$^2$.
All peaks appear at small-gap regions, similar to nonlinear Hall conductivities in e.g. MnS \cite{Wang_2023} and nonlinear spin generations in e.g. MnBi$_2$Te$_4$ \cite{Xiao_2022}.

Fig. \ref{fig04}b demonstrates $\chi^{\text{o}}_{zxx,A}$ and $\chi^{\text{o}}_{zxx,B}$ as functions of the canting angle, $\theta$, with fixed $\phi =0$.
The out-of-plane spin generations on the two Fe atoms are staggered and opposite to local magnetic moments. Moreover, their variations with $\theta$ are odd with respect to $\theta=0$. These first-principles based results all support the model calculations in Fig. \ref{fig02}b.

In addition, staggered and out-of-plane nonlinear spin generations have generalizations to other altermagnets, e.g. V$_2$Se$_2$O monolayer, which has been also studied by first-principles calculations and provided in SI.

{\color{blue}\textit{Conclusion}} --
To conclude, by symmetry analyses, tight-binding model and first-principles calculations, we investigate nonlinear spin generations in centrosymmetric altermagnets.
Staggered and out-of-plane spin generations are found in both a square altermagnetic model and a typical altermagnet, Fe$_2$Se$_2$O monolayer.
These nonlinear spin generations pave avenues for manipulating magnetic dynamics by electric means in compensated magnets with opposite-spin sublattices, and realizing perpendicular magnetization switching in high-density storage devices.

{\color{blue}\textit{Acknowledgments}} -- 
We appreciate helpful discussions with Huiying Liu. 
We are supported by the National Natural Science Foundation of China (Nos. 12374044, 12004186, 11904173). 

\appendix



\begin{thebibliography}{30}%
\makeatletter
\providecommand \@ifxundefined [1]{%
 \@ifx{#1\undefined}
}%
\providecommand \@ifnum [1]{%
 \ifnum #1\expandafter \@firstoftwo
 \else \expandafter \@secondoftwo
 \fi
}%
\providecommand \@ifx [1]{%
 \ifx #1\expandafter \@firstoftwo
 \else \expandafter \@secondoftwo
 \fi
}%
\providecommand \natexlab [1]{#1}%
\providecommand \enquote  [1]{``#1''}%
\providecommand \bibnamefont  [1]{#1}%
\providecommand \bibfnamefont [1]{#1}%
\providecommand \citenamefont [1]{#1}%
\providecommand \href@noop [0]{\@secondoftwo}%
\providecommand \href [0]{\begingroup \@sanitize@url \@href}%
\providecommand \@href[1]{\@@startlink{#1}\@@href}%
\providecommand \@@href[1]{\endgroup#1\@@endlink}%
\providecommand \@sanitize@url [0]{\catcode `\\12\catcode `\$12\catcode `\&12\catcode `\#12\catcode `\^12\catcode `\_12\catcode `\%12\relax}%
\providecommand \@@startlink[1]{}%
\providecommand \@@endlink[0]{}%
\providecommand \url  [0]{\begingroup\@sanitize@url \@url }%
\providecommand \@url [1]{\endgroup\@href {#1}{\urlprefix }}%
\providecommand \urlprefix  [0]{URL }%
\providecommand \Eprint [0]{\href }%
\providecommand \doibase [0]{http://dx.doi.org/}%
\providecommand \selectlanguage [0]{\@gobble}%
\providecommand \bibinfo  [0]{\@secondoftwo}%
\providecommand \bibfield  [0]{\@secondoftwo}%
\providecommand \translation [1]{[#1]}%
\providecommand \BibitemOpen [0]{}%
\providecommand \bibitemStop [0]{}%
\providecommand \bibitemNoStop [0]{.\EOS\space}%
\providecommand \EOS [0]{\spacefactor3000\relax}%
\providecommand \BibitemShut  [1]{\csname bibitem#1\endcsname}%
\let\auto@bib@innerbib\@empty
\bibitem [{\citenamefont {Manchon}\ \emph {et~al.}(2019)\citenamefont {Manchon}, \citenamefont {\ifmmode~\check{Z}\else \v{Z}\fi{}elezn\'y}, \citenamefont {Miron}, \citenamefont {Jungwirth}, \citenamefont {Sinova}, \citenamefont {Thiaville}, \citenamefont {Garello},\ and\ \citenamefont {Gambardella}}]{RevModPhys.91.035004}%
  \BibitemOpen
  \bibfield  {author} {\bibinfo {author} {\bibfnamefont {A.}~\bibnamefont {Manchon}}, \bibinfo {author} {\bibfnamefont {J.}~\bibnamefont {\ifmmode~\check{Z}\else \v{Z}\fi{}elezn\'y}}, \bibinfo {author} {\bibfnamefont {I.~M.}\ \bibnamefont {Miron}}, \bibinfo {author} {\bibfnamefont {T.}~\bibnamefont {Jungwirth}}, \bibinfo {author} {\bibfnamefont {J.}~\bibnamefont {Sinova}}, \bibinfo {author} {\bibfnamefont {A.}~\bibnamefont {Thiaville}}, \bibinfo {author} {\bibfnamefont {K.}~\bibnamefont {Garello}}, \ and\ \bibinfo {author} {\bibfnamefont {P.}~\bibnamefont {Gambardella}},\ }\href {\doibase 10.1103/RevModPhys.91.035004} {\bibfield  {journal} {\bibinfo  {journal} {Rev. Mod. Phys.}\ }\textbf {\bibinfo {volume} {91}},\ \bibinfo {pages} {035004} (\bibinfo {year} {2019})}\BibitemShut {NoStop}%
\bibitem [{\citenamefont {Chappert}\ \emph {et~al.}(2007)\citenamefont {Chappert}, \citenamefont {Fert},\ and\ \citenamefont {Van~Dau}}]{Chappert_Fert_and_Dau_2007}%
  \BibitemOpen
  \bibfield  {author} {\bibinfo {author} {\bibfnamefont {C.}~\bibnamefont {Chappert}}, \bibinfo {author} {\bibfnamefont {A.}~\bibnamefont {Fert}}, \ and\ \bibinfo {author} {\bibfnamefont {F.~N.}\ \bibnamefont {Van~Dau}},\ }\href {\doibase 10.1038/nmat2024} {\bibfield  {journal} {\bibinfo  {journal} {Nat. Mater.}\ }\textbf {\bibinfo {volume} {6}},\ \bibinfo {pages} {813} (\bibinfo {year} {2007})}\BibitemShut {NoStop}%
\bibitem [{\citenamefont {Brataas}\ \emph {et~al.}(2012)\citenamefont {Brataas}, \citenamefont {Kent},\ and\ \citenamefont {Ohno}}]{Brataas_Kent_and_Ohno_2012}%
  \BibitemOpen
  \bibfield  {author} {\bibinfo {author} {\bibfnamefont {A.}~\bibnamefont {Brataas}}, \bibinfo {author} {\bibfnamefont {A.~D.}\ \bibnamefont {Kent}}, \ and\ \bibinfo {author} {\bibfnamefont {H.}~\bibnamefont {Ohno}},\ }\href {\doibase 10.1038/NMAT3311} {\bibfield  {journal} {\bibinfo  {journal} {Nat. Mater.}\ }\textbf {\bibinfo {volume} {11}},\ \bibinfo {pages} {372} (\bibinfo {year} {2012})}\BibitemShut {NoStop}%
\bibitem [{\citenamefont {\ifmmode~\check{Z}\else \v{Z}\fi{}elezn\'y}\ \emph {et~al.}(2014)\citenamefont {\ifmmode~\check{Z}\else \v{Z}\fi{}elezn\'y}, \citenamefont {Gao}, \citenamefont {V\'yborn\'y}, \citenamefont {Zemen}, \citenamefont {Ma\ifmmode~\check{s}\else \v{s}\fi{}ek}, \citenamefont {Manchon}, \citenamefont {Wunderlich}, \citenamefont {Sinova},\ and\ \citenamefont {Jungwirth}}]{Zelezny_2014}%
  \BibitemOpen
  \bibfield  {author} {\bibinfo {author} {\bibfnamefont {J.}~\bibnamefont {\ifmmode~\check{Z}\else \v{Z}\fi{}elezn\'y}}, \bibinfo {author} {\bibfnamefont {H.}~\bibnamefont {Gao}}, \bibinfo {author} {\bibfnamefont {K.}~\bibnamefont {V\'yborn\'y}}, \bibinfo {author} {\bibfnamefont {J.}~\bibnamefont {Zemen}}, \bibinfo {author} {\bibfnamefont {J.}~\bibnamefont {Ma\ifmmode~\check{s}\else \v{s}\fi{}ek}}, \bibinfo {author} {\bibfnamefont {A.}~\bibnamefont {Manchon}}, \bibinfo {author} {\bibfnamefont {J.}~\bibnamefont {Wunderlich}}, \bibinfo {author} {\bibfnamefont {J.}~\bibnamefont {Sinova}}, \ and\ \bibinfo {author} {\bibfnamefont {T.}~\bibnamefont {Jungwirth}},\ }\href {\doibase 10.1103/PhysRevLett.113.157201} {\bibfield  {journal} {\bibinfo  {journal} {Phys. Rev. Lett.}\ }\textbf {\bibinfo {volume} {113}},\ \bibinfo {pages} {157201} (\bibinfo {year} {2014})}\BibitemShut {NoStop}%
\bibitem [{\citenamefont {\ifmmode~\check{Z}\else \v{Z}\fi{}elezn\'y}\ \emph {et~al.}(2017)\citenamefont {\ifmmode~\check{Z}\else \v{Z}\fi{}elezn\'y}, \citenamefont {Gao}, \citenamefont {Manchon}, \citenamefont {Freimuth}, \citenamefont {Mokrousov}, \citenamefont {Zemen}, \citenamefont {Ma\ifmmode~\check{s}\else \v{s}\fi{}ek}, \citenamefont {Sinova},\ and\ \citenamefont {Jungwirth}}]{Zelezny_2017}%
  \BibitemOpen
  \bibfield  {author} {\bibinfo {author} {\bibfnamefont {J.}~\bibnamefont {\ifmmode~\check{Z}\else \v{Z}\fi{}elezn\'y}}, \bibinfo {author} {\bibfnamefont {H.}~\bibnamefont {Gao}}, \bibinfo {author} {\bibfnamefont {A.}~\bibnamefont {Manchon}}, \bibinfo {author} {\bibfnamefont {F.}~\bibnamefont {Freimuth}}, \bibinfo {author} {\bibfnamefont {Y.}~\bibnamefont {Mokrousov}}, \bibinfo {author} {\bibfnamefont {J.}~\bibnamefont {Zemen}}, \bibinfo {author} {\bibfnamefont {J.}~\bibnamefont {Ma\ifmmode~\check{s}\else \v{s}\fi{}ek}}, \bibinfo {author} {\bibfnamefont {J.}~\bibnamefont {Sinova}}, \ and\ \bibinfo {author} {\bibfnamefont {T.}~\bibnamefont {Jungwirth}},\ }\href {\doibase 10.1103/PhysRevB.95.014403} {\bibfield  {journal} {\bibinfo  {journal} {Phys. Rev. B}\ }\textbf {\bibinfo {volume} {95}},\ \bibinfo {pages} {014403} (\bibinfo {year} {2017})}\BibitemShut {NoStop}%
\bibitem [{\citenamefont {Wadley}\ \emph {et~al.}(2016)\citenamefont {Wadley}, \citenamefont {Howells}, \citenamefont {Železný}, \citenamefont {Andrews}, \citenamefont {Hills}, \citenamefont {Campion}, \citenamefont {Novák}, \citenamefont {Olejník}, \citenamefont {Maccherozzi}, \citenamefont {Dhesi}, \citenamefont {Martin}, \citenamefont {Wagner}, \citenamefont {Wunderlich}, \citenamefont {Freimuth}, \citenamefont {Mokrousov}, \citenamefont {Kuneš}, \citenamefont {Chauhan}, \citenamefont {Grzybowski}, \citenamefont {Rushforth}, \citenamefont {Edmonds}, \citenamefont {Gallagher},\ and\ \citenamefont {Jungwirth}}]{Wadley_2016}%
  \BibitemOpen
  \bibfield  {author} {\bibinfo {author} {\bibfnamefont {P.}~\bibnamefont {Wadley}}, \bibinfo {author} {\bibfnamefont {B.}~\bibnamefont {Howells}}, \bibinfo {author} {\bibfnamefont {J.}~\bibnamefont {Železný}}, \bibinfo {author} {\bibfnamefont {C.}~\bibnamefont {Andrews}}, \bibinfo {author} {\bibfnamefont {V.}~\bibnamefont {Hills}}, \bibinfo {author} {\bibfnamefont {R.~P.}\ \bibnamefont {Campion}}, \bibinfo {author} {\bibfnamefont {V.}~\bibnamefont {Novák}}, \bibinfo {author} {\bibfnamefont {K.}~\bibnamefont {Olejník}}, \bibinfo {author} {\bibfnamefont {F.}~\bibnamefont {Maccherozzi}}, \bibinfo {author} {\bibfnamefont {S.~S.}\ \bibnamefont {Dhesi}}, \bibinfo {author} {\bibfnamefont {S.~Y.}\ \bibnamefont {Martin}}, \bibinfo {author} {\bibfnamefont {T.}~\bibnamefont {Wagner}}, \bibinfo {author} {\bibfnamefont {J.}~\bibnamefont {Wunderlich}}, \bibinfo {author} {\bibfnamefont {F.}~\bibnamefont {Freimuth}}, \bibinfo {author} {\bibfnamefont {Y.}~\bibnamefont {Mokrousov}}, \bibinfo {author} {\bibfnamefont {J.}~\bibnamefont {Kuneš}}, \bibinfo {author} {\bibfnamefont {J.~S.}\ \bibnamefont {Chauhan}}, \bibinfo {author} {\bibfnamefont {M.~J.}\ \bibnamefont {Grzybowski}}, \bibinfo {author} {\bibfnamefont {A.~W.}\ \bibnamefont {Rushforth}}, \bibinfo {author} {\bibfnamefont {K.~W.}\ \bibnamefont {Edmonds}}, \bibinfo {author} {\bibfnamefont {B.~L.}\ \bibnamefont {Gallagher}}, \ and\ \bibinfo {author} {\bibfnamefont {T.}~\bibnamefont {Jungwirth}},\ }\href {\doibase 10.1126/science.aab1031} {\bibfield  {journal} {\bibinfo  {journal} {Science}\ }\textbf {\bibinfo {volume} {351}},\ \bibinfo {pages} {587} (\bibinfo {year} {2016})}\BibitemShut {NoStop}%
\bibitem [{\citenamefont {Bodnar}\ \emph {et~al.}(2018)\citenamefont {Bodnar}, \citenamefont {Smejkal}, \citenamefont {Turek}, \citenamefont {Jungwirth}, \citenamefont {Gomonay}, \citenamefont {Sinova}, \citenamefont {Sapozhnik}, \citenamefont {Elmers}, \citenamefont {Klaui},\ and\ \citenamefont {Jourdan}}]{Bodnar_Smejkal_2018}%
  \BibitemOpen
  \bibfield  {author} {\bibinfo {author} {\bibfnamefont {S.~Y.}\ \bibnamefont {Bodnar}}, \bibinfo {author} {\bibfnamefont {L.}~\bibnamefont {Smejkal}}, \bibinfo {author} {\bibfnamefont {I.}~\bibnamefont {Turek}}, \bibinfo {author} {\bibfnamefont {T.}~\bibnamefont {Jungwirth}}, \bibinfo {author} {\bibfnamefont {O.}~\bibnamefont {Gomonay}}, \bibinfo {author} {\bibfnamefont {J.}~\bibnamefont {Sinova}}, \bibinfo {author} {\bibfnamefont {A.~A.}\ \bibnamefont {Sapozhnik}}, \bibinfo {author} {\bibfnamefont {H.~J.}\ \bibnamefont {Elmers}}, \bibinfo {author} {\bibfnamefont {M.}~\bibnamefont {Klaui}}, \ and\ \bibinfo {author} {\bibfnamefont {M.}~\bibnamefont {Jourdan}},\ }\href {\doibase 10.1038/s41467-017-02780-x} {\bibfield  {journal} {\bibinfo  {journal} {Nat. Commun.}\ }\textbf {\bibinfo {volume} {9}},\ \bibinfo {pages} {348} (\bibinfo {year} {2018})}\BibitemShut {NoStop}%
\bibitem [{\citenamefont {Zhou}\ \emph {et~al.}(2018)\citenamefont {Zhou}, \citenamefont {Zhang}, \citenamefont {Li}, \citenamefont {Chen}, \citenamefont {Shi}, \citenamefont {Tan}, \citenamefont {Gu}, \citenamefont {Saleem}, \citenamefont {Wu}, \citenamefont {Pan},\ and\ \citenamefont {Song}}]{Zhou_Zhang_li_2018}%
  \BibitemOpen
  \bibfield  {author} {\bibinfo {author} {\bibfnamefont {X.~F.}\ \bibnamefont {Zhou}}, \bibinfo {author} {\bibfnamefont {J.}~\bibnamefont {Zhang}}, \bibinfo {author} {\bibfnamefont {F.}~\bibnamefont {Li}}, \bibinfo {author} {\bibfnamefont {X.~Z.}\ \bibnamefont {Chen}}, \bibinfo {author} {\bibfnamefont {G.~Y.}\ \bibnamefont {Shi}}, \bibinfo {author} {\bibfnamefont {Y.~Z.}\ \bibnamefont {Tan}}, \bibinfo {author} {\bibfnamefont {Y.~D.}\ \bibnamefont {Gu}}, \bibinfo {author} {\bibfnamefont {M.~S.}\ \bibnamefont {Saleem}}, \bibinfo {author} {\bibfnamefont {H.~Q.}\ \bibnamefont {Wu}}, \bibinfo {author} {\bibfnamefont {F.}~\bibnamefont {Pan}}, \ and\ \bibinfo {author} {\bibfnamefont {C.}~\bibnamefont {Song}},\ }\href {\doibase 10.1103/PhysRevApplied.9.054028} {\bibfield  {journal} {\bibinfo  {journal} {Phys. Rev. Appl.}\ }\textbf {\bibinfo {volume} {9}},\ \bibinfo {pages} {054028} (\bibinfo {year} {2018})}\BibitemShut {NoStop}%
\bibitem [{\citenamefont {Selzer}\ \emph {et~al.}(2022)\citenamefont {Selzer}, \citenamefont {Salemi}, \citenamefont {De\'ak}, \citenamefont {Simon}, \citenamefont {Szunyogh}, \citenamefont {Oppeneer},\ and\ \citenamefont {Nowak}}]{Selzer_Salemi_Deak_2022}%
  \BibitemOpen
  \bibfield  {author} {\bibinfo {author} {\bibfnamefont {S.}~\bibnamefont {Selzer}}, \bibinfo {author} {\bibfnamefont {L.}~\bibnamefont {Salemi}}, \bibinfo {author} {\bibfnamefont {A.}~\bibnamefont {De\'ak}}, \bibinfo {author} {\bibfnamefont {E.}~\bibnamefont {Simon}}, \bibinfo {author} {\bibfnamefont {L.}~\bibnamefont {Szunyogh}}, \bibinfo {author} {\bibfnamefont {P.~M.}\ \bibnamefont {Oppeneer}}, \ and\ \bibinfo {author} {\bibfnamefont {U.}~\bibnamefont {Nowak}},\ }\href {\doibase 10.1103/PhysRevB.105.174416} {\bibfield  {journal} {\bibinfo  {journal} {Phys. Rev. B}\ }\textbf {\bibinfo {volume} {105}},\ \bibinfo {pages} {174416} (\bibinfo {year} {2022})}\BibitemShut {NoStop}%
\bibitem [{\citenamefont {Xiao}\ \emph {et~al.}(2022)\citenamefont {Xiao}, \citenamefont {Liu}, \citenamefont {Wu}, \citenamefont {Wang}, \citenamefont {Niu},\ and\ \citenamefont {Yang}}]{Xiao_2022}%
  \BibitemOpen
  \bibfield  {author} {\bibinfo {author} {\bibfnamefont {C.}~\bibnamefont {Xiao}}, \bibinfo {author} {\bibfnamefont {H.}~\bibnamefont {Liu}}, \bibinfo {author} {\bibfnamefont {W.}~\bibnamefont {Wu}}, \bibinfo {author} {\bibfnamefont {H.}~\bibnamefont {Wang}}, \bibinfo {author} {\bibfnamefont {Q.}~\bibnamefont {Niu}}, \ and\ \bibinfo {author} {\bibfnamefont {S.~A.}\ \bibnamefont {Yang}},\ }\href {\doibase 10.1103/PhysRevLett.129.086602} {\bibfield  {journal} {\bibinfo  {journal} {Phys. Rev. Lett.}\ }\textbf {\bibinfo {volume} {129}},\ \bibinfo {pages} {086602} (\bibinfo {year} {2022})}\BibitemShut {NoStop}%
\bibitem [{\citenamefont {Xiao}\ \emph {et~al.}(2023)\citenamefont {Xiao}, \citenamefont {Wu}, \citenamefont {Wang}, \citenamefont {Huang}, \citenamefont {Feng}, \citenamefont {Liu}, \citenamefont {Guo}, \citenamefont {Niu},\ and\ \citenamefont {Yang}}]{Xiao_2023}%
  \BibitemOpen
  \bibfield  {author} {\bibinfo {author} {\bibfnamefont {C.}~\bibnamefont {Xiao}}, \bibinfo {author} {\bibfnamefont {W.}~\bibnamefont {Wu}}, \bibinfo {author} {\bibfnamefont {H.}~\bibnamefont {Wang}}, \bibinfo {author} {\bibfnamefont {Y.-X.}\ \bibnamefont {Huang}}, \bibinfo {author} {\bibfnamefont {X.}~\bibnamefont {Feng}}, \bibinfo {author} {\bibfnamefont {H.}~\bibnamefont {Liu}}, \bibinfo {author} {\bibfnamefont {G.-Y.}\ \bibnamefont {Guo}}, \bibinfo {author} {\bibfnamefont {Q.}~\bibnamefont {Niu}}, \ and\ \bibinfo {author} {\bibfnamefont {S.~A.}\ \bibnamefont {Yang}},\ }\href {\doibase 10.1103/PhysRevLett.130.166302} {\bibfield  {journal} {\bibinfo  {journal} {Phys. Rev. Lett.}\ }\textbf {\bibinfo {volume} {130}},\ \bibinfo {pages} {166302} (\bibinfo {year} {2023})}\BibitemShut {NoStop}%
\bibitem [{\citenamefont {Hayami}\ \emph {et~al.}(2019)\citenamefont {Hayami}, \citenamefont {Yanagi},\ and\ \citenamefont {Kusunose}}]{Hayami_2019}%
  \BibitemOpen
  \bibfield  {author} {\bibinfo {author} {\bibfnamefont {S.}~\bibnamefont {Hayami}}, \bibinfo {author} {\bibfnamefont {Y.}~\bibnamefont {Yanagi}}, \ and\ \bibinfo {author} {\bibfnamefont {H.}~\bibnamefont {Kusunose}},\ }\href {\doibase 10.7566/JPSJ.88.123702} {\bibfield  {journal} {\bibinfo  {journal} {J. Phys. Soc. Jpn}\ }\textbf {\bibinfo {volume} {88}},\ \bibinfo {pages} {123702} (\bibinfo {year} {2019})}\BibitemShut {NoStop}%
\bibitem [{\citenamefont {Yuan}\ \emph {et~al.}(2020)\citenamefont {Yuan}, \citenamefont {Wang}, \citenamefont {Luo}, \citenamefont {Rashba},\ and\ \citenamefont {Zunger}}]{Yuan_2020}%
  \BibitemOpen
  \bibfield  {author} {\bibinfo {author} {\bibfnamefont {L.-D.}\ \bibnamefont {Yuan}}, \bibinfo {author} {\bibfnamefont {Z.}~\bibnamefont {Wang}}, \bibinfo {author} {\bibfnamefont {J.-W.}\ \bibnamefont {Luo}}, \bibinfo {author} {\bibfnamefont {E.~I.}\ \bibnamefont {Rashba}}, \ and\ \bibinfo {author} {\bibfnamefont {A.}~\bibnamefont {Zunger}},\ }\href {\doibase 10.1103/PhysRevB.102.014422} {\bibfield  {journal} {\bibinfo  {journal} {Phys. Rev. B}\ }\textbf {\bibinfo {volume} {102}},\ \bibinfo {pages} {014422} (\bibinfo {year} {2020})}\BibitemShut {NoStop}%
\bibitem [{\citenamefont {Hayami}\ \emph {et~al.}(2020)\citenamefont {Hayami}, \citenamefont {Yanagi},\ and\ \citenamefont {Kusunose}}]{Hayami_2020}%
  \BibitemOpen
  \bibfield  {author} {\bibinfo {author} {\bibfnamefont {S.}~\bibnamefont {Hayami}}, \bibinfo {author} {\bibfnamefont {Y.}~\bibnamefont {Yanagi}}, \ and\ \bibinfo {author} {\bibfnamefont {H.}~\bibnamefont {Kusunose}},\ }\href {\doibase 10.1103/PhysRevB.102.144441} {\bibfield  {journal} {\bibinfo  {journal} {Phys. Rev. B}\ }\textbf {\bibinfo {volume} {102}},\ \bibinfo {pages} {144441} (\bibinfo {year} {2020})}\BibitemShut {NoStop}%
\bibitem [{\citenamefont {Liu}\ \emph {et~al.}(2022)\citenamefont {Liu}, \citenamefont {Li}, \citenamefont {Han}, \citenamefont {Wan},\ and\ \citenamefont {Liu}}]{Liu_2022}%
  \BibitemOpen
  \bibfield  {author} {\bibinfo {author} {\bibfnamefont {P.}~\bibnamefont {Liu}}, \bibinfo {author} {\bibfnamefont {J.}~\bibnamefont {Li}}, \bibinfo {author} {\bibfnamefont {J.}~\bibnamefont {Han}}, \bibinfo {author} {\bibfnamefont {X.}~\bibnamefont {Wan}}, \ and\ \bibinfo {author} {\bibfnamefont {Q.}~\bibnamefont {Liu}},\ }\href {\doibase 10.1103/PhysRevX.12.021016} {\bibfield  {journal} {\bibinfo  {journal} {Phys. Rev. X}\ }\textbf {\bibinfo {volume} {12}},\ \bibinfo {pages} {021016} (\bibinfo {year} {2022})}\BibitemShut {NoStop}%
\bibitem [{\citenamefont {\ifmmode~\check{S}\else \v{S}\fi{}mejkal}\ \emph {et~al.}(2022{\natexlab{a}})\citenamefont {\ifmmode~\check{S}\else \v{S}\fi{}mejkal}, \citenamefont {Sinova},\ and\ \citenamefont {Jungwirth}}]{Smejkal_2022}%
  \BibitemOpen
  \bibfield  {author} {\bibinfo {author} {\bibfnamefont {L.}~\bibnamefont {\ifmmode~\check{S}\else \v{S}\fi{}mejkal}}, \bibinfo {author} {\bibfnamefont {J.}~\bibnamefont {Sinova}}, \ and\ \bibinfo {author} {\bibfnamefont {T.}~\bibnamefont {Jungwirth}},\ }\href {\doibase 10.1103/PhysRevX.12.031042} {\bibfield  {journal} {\bibinfo  {journal} {Phys. Rev. X}\ }\textbf {\bibinfo {volume} {12}},\ \bibinfo {pages} {031042} (\bibinfo {year} {2022}{\natexlab{a}})}\BibitemShut {NoStop}%
\bibitem [{\citenamefont {\ifmmode~\check{S}\else \v{S}\fi{}mejkal}\ \emph {et~al.}(2022{\natexlab{b}})\citenamefont {\ifmmode~\check{S}\else \v{S}\fi{}mejkal}, \citenamefont {Sinova},\ and\ \citenamefont {Jungwirth}}]{Smejkal_Sinova_2022}%
  \BibitemOpen
  \bibfield  {author} {\bibinfo {author} {\bibfnamefont {L.}~\bibnamefont {\ifmmode~\check{S}\else \v{S}\fi{}mejkal}}, \bibinfo {author} {\bibfnamefont {J.}~\bibnamefont {Sinova}}, \ and\ \bibinfo {author} {\bibfnamefont {T.}~\bibnamefont {Jungwirth}},\ }\href {\doibase 10.1103/PhysRevX.12.040501} {\bibfield  {journal} {\bibinfo  {journal} {Phys. Rev. X}\ }\textbf {\bibinfo {volume} {12}},\ \bibinfo {pages} {040501} (\bibinfo {year} {2022}{\natexlab{b}})}\BibitemShut {NoStop}%
\bibitem [{\citenamefont {Han}\ \emph {et~al.}(2024)\citenamefont {Han}, \citenamefont {Fu}, \citenamefont {Peng}, \citenamefont {Cheng}, \citenamefont {Dai}, \citenamefont {Liu}, \citenamefont {Li}, \citenamefont {Zhang}, \citenamefont {Zhu}, \citenamefont {Bai}, \citenamefont {Zhou}, \citenamefont {Liang}, \citenamefont {Chen}, \citenamefont {Wang}, \citenamefont {Chen}, \citenamefont {Yang}, \citenamefont {Zhang}, \citenamefont {Song}, \citenamefont {Liu},\ and\ \citenamefont {Pan}}]{Han_2024}%
  \BibitemOpen
  \bibfield  {author} {\bibinfo {author} {\bibfnamefont {L.}~\bibnamefont {Han}}, \bibinfo {author} {\bibfnamefont {X.}~\bibnamefont {Fu}}, \bibinfo {author} {\bibfnamefont {R.}~\bibnamefont {Peng}}, \bibinfo {author} {\bibfnamefont {X.}~\bibnamefont {Cheng}}, \bibinfo {author} {\bibfnamefont {J.}~\bibnamefont {Dai}}, \bibinfo {author} {\bibfnamefont {L.}~\bibnamefont {Liu}}, \bibinfo {author} {\bibfnamefont {Y.}~\bibnamefont {Li}}, \bibinfo {author} {\bibfnamefont {Y.}~\bibnamefont {Zhang}}, \bibinfo {author} {\bibfnamefont {W.}~\bibnamefont {Zhu}}, \bibinfo {author} {\bibfnamefont {H.}~\bibnamefont {Bai}}, \bibinfo {author} {\bibfnamefont {Y.}~\bibnamefont {Zhou}}, \bibinfo {author} {\bibfnamefont {S.}~\bibnamefont {Liang}}, \bibinfo {author} {\bibfnamefont {C.}~\bibnamefont {Chen}}, \bibinfo {author} {\bibfnamefont {Q.}~\bibnamefont {Wang}}, \bibinfo {author} {\bibfnamefont {X.}~\bibnamefont {Chen}}, \bibinfo {author} {\bibfnamefont {L.}~\bibnamefont {Yang}}, \bibinfo {author} {\bibfnamefont {Y.}~\bibnamefont {Zhang}}, \bibinfo {author} {\bibfnamefont {C.}~\bibnamefont {Song}}, \bibinfo {author} {\bibfnamefont {J.}~\bibnamefont {Liu}}, \ and\ \bibinfo {author} {\bibfnamefont {F.}~\bibnamefont {Pan}},\ }\href {\doibase 10.1126/sciadv.adn0479} {\bibfield  {journal} {\bibinfo  {journal} {Sci. Adv.}\ }\textbf {\bibinfo {volume} {10}},\ \bibinfo {pages} {eadn0479} (\bibinfo {year} {2024})}\BibitemShut {NoStop}%
\bibitem [{\citenamefont {Urata}\ \emph {et~al.}(2024)\citenamefont {Urata}, \citenamefont {Hattori},\ and\ \citenamefont {Ikuta}}]{Urata_2024}%
  \BibitemOpen
  \bibfield  {author} {\bibinfo {author} {\bibfnamefont {T.}~\bibnamefont {Urata}}, \bibinfo {author} {\bibfnamefont {W.}~\bibnamefont {Hattori}}, \ and\ \bibinfo {author} {\bibfnamefont {H.}~\bibnamefont {Ikuta}},\ }\href {\doibase 10.1103/PhysRevMaterials.8.084412} {\bibfield  {journal} {\bibinfo  {journal} {Phys. Rev. Mater.}\ }\textbf {\bibinfo {volume} {8}},\ \bibinfo {pages} {084412} (\bibinfo {year} {2024})}\BibitemShut {NoStop}%
\bibitem [{\citenamefont {Brekke}\ \emph {et~al.}(2023)\citenamefont {Brekke}, \citenamefont {Brataas},\ and\ \citenamefont {Sudb\o{}}}]{Brekke_2023}%
  \BibitemOpen
  \bibfield  {author} {\bibinfo {author} {\bibfnamefont {B.}~\bibnamefont {Brekke}}, \bibinfo {author} {\bibfnamefont {A.}~\bibnamefont {Brataas}}, \ and\ \bibinfo {author} {\bibfnamefont {A.}~\bibnamefont {Sudb\o{}}},\ }\href {\doibase 10.1103/PhysRevB.108.224421} {\bibfield  {journal} {\bibinfo  {journal} {Phys. Rev. B}\ }\textbf {\bibinfo {volume} {108}},\ \bibinfo {pages} {224421} (\bibinfo {year} {2023})}\BibitemShut {NoStop}%
\bibitem [{\citenamefont {Orozco-Galvan}\ \emph {et~al.}(2024)\citenamefont {Orozco-Galvan}, \citenamefont {Garc\'{\i}a-Fuente},\ and\ \citenamefont {Barraza-Lopez}}]{Gustavo_2024}%
  \BibitemOpen
  \bibfield  {author} {\bibinfo {author} {\bibfnamefont {G.~S.}\ \bibnamefont {Orozco-Galvan}}, \bibinfo {author} {\bibfnamefont {A.}~\bibnamefont {Garc\'{\i}a-Fuente}}, \ and\ \bibinfo {author} {\bibfnamefont {S.}~\bibnamefont {Barraza-Lopez}},\ }\href {\doibase 10.1103/PhysRevB.109.035141} {\bibfield  {journal} {\bibinfo  {journal} {Phys. Rev. B}\ }\textbf {\bibinfo {volume} {109}},\ \bibinfo {pages} {035141} (\bibinfo {year} {2024})}\BibitemShut {NoStop}%
\bibitem [{\citenamefont {Attias}\ \emph {et~al.}(2024)\citenamefont {Attias}, \citenamefont {Levchenko},\ and\ \citenamefont {Khodas}}]{Attias_2024}%
  \BibitemOpen
  \bibfield  {author} {\bibinfo {author} {\bibfnamefont {L.}~\bibnamefont {Attias}}, \bibinfo {author} {\bibfnamefont {A.}~\bibnamefont {Levchenko}}, \ and\ \bibinfo {author} {\bibfnamefont {M.}~\bibnamefont {Khodas}},\ }\href {\doibase 10.1103/PhysRevB.110.094425} {\bibfield  {journal} {\bibinfo  {journal} {Phys. Rev. B}\ }\textbf {\bibinfo {volume} {110}},\ \bibinfo {pages} {094425} (\bibinfo {year} {2024})}\BibitemShut {NoStop}%
\bibitem [{\citenamefont {Shao}\ \emph {et~al.}(2023)\citenamefont {Shao}, \citenamefont {Jiang}, \citenamefont {Ding}, \citenamefont {Zhang}, \citenamefont {Wang}, \citenamefont {Xiao}, \citenamefont {Gurung}, \citenamefont {Lu}, \citenamefont {Sun},\ and\ \citenamefont {Tsymbal}}]{Shao_2023}%
  \BibitemOpen
  \bibfield  {author} {\bibinfo {author} {\bibfnamefont {D.-F.}\ \bibnamefont {Shao}}, \bibinfo {author} {\bibfnamefont {Y.-Y.}\ \bibnamefont {Jiang}}, \bibinfo {author} {\bibfnamefont {J.}~\bibnamefont {Ding}}, \bibinfo {author} {\bibfnamefont {S.-H.}\ \bibnamefont {Zhang}}, \bibinfo {author} {\bibfnamefont {Z.-A.}\ \bibnamefont {Wang}}, \bibinfo {author} {\bibfnamefont {R.-C.}\ \bibnamefont {Xiao}}, \bibinfo {author} {\bibfnamefont {G.}~\bibnamefont {Gurung}}, \bibinfo {author} {\bibfnamefont {W.~J.}\ \bibnamefont {Lu}}, \bibinfo {author} {\bibfnamefont {Y.~P.}\ \bibnamefont {Sun}}, \ and\ \bibinfo {author} {\bibfnamefont {E.~Y.}\ \bibnamefont {Tsymbal}},\ }\href {\doibase 10.1103/PhysRevLett.130.216702} {\bibfield  {journal} {\bibinfo  {journal} {Phys. Rev. Lett.}\ }\textbf {\bibinfo {volume} {130}},\ \bibinfo {pages} {216702} (\bibinfo {year} {2023})}\BibitemShut {NoStop}%
\bibitem [{\citenamefont {Zhang}\ \emph {et~al.}(2022)\citenamefont {Zhang}, \citenamefont {Chou}, \citenamefont {Yun}, \citenamefont {McGoldrick}, \citenamefont {Hou}, \citenamefont {Mkhoyan},\ and\ \citenamefont {Liu}}]{Zhang_2022}%
  \BibitemOpen
  \bibfield  {author} {\bibinfo {author} {\bibfnamefont {P.}~\bibnamefont {Zhang}}, \bibinfo {author} {\bibfnamefont {C.-T.}\ \bibnamefont {Chou}}, \bibinfo {author} {\bibfnamefont {H.}~\bibnamefont {Yun}}, \bibinfo {author} {\bibfnamefont {B.~C.}\ \bibnamefont {McGoldrick}}, \bibinfo {author} {\bibfnamefont {J.~T.}\ \bibnamefont {Hou}}, \bibinfo {author} {\bibfnamefont {K.~A.}\ \bibnamefont {Mkhoyan}}, \ and\ \bibinfo {author} {\bibfnamefont {L.}~\bibnamefont {Liu}},\ }\href {\doibase 10.1103/PhysRevLett.129.017203} {\bibfield  {journal} {\bibinfo  {journal} {Phys. Rev. Lett.}\ }\textbf {\bibinfo {volume} {129}},\ \bibinfo {pages} {017203} (\bibinfo {year} {2022})}\BibitemShut {NoStop}%
\bibitem [{\citenamefont {Hu}\ \emph {et~al.}(2022)\citenamefont {Hu}, \citenamefont {Shao}, \citenamefont {Yang}, \citenamefont {Pan}, \citenamefont {Fu}, \citenamefont {Tang}, \citenamefont {Yang}, \citenamefont {Fan}, \citenamefont {Zhou}, \citenamefont {Tsymbal},\ and\ \citenamefont {Qiu}}]{Shuai_2022}%
  \BibitemOpen
  \bibfield  {author} {\bibinfo {author} {\bibfnamefont {S.}~\bibnamefont {Hu}}, \bibinfo {author} {\bibfnamefont {D.-F.}\ \bibnamefont {Shao}}, \bibinfo {author} {\bibfnamefont {H.}~\bibnamefont {Yang}}, \bibinfo {author} {\bibfnamefont {C.}~\bibnamefont {Pan}}, \bibinfo {author} {\bibfnamefont {Z.}~\bibnamefont {Fu}}, \bibinfo {author} {\bibfnamefont {M.}~\bibnamefont {Tang}}, \bibinfo {author} {\bibfnamefont {Y.}~\bibnamefont {Yang}}, \bibinfo {author} {\bibfnamefont {W.}~\bibnamefont {Fan}}, \bibinfo {author} {\bibfnamefont {S.}~\bibnamefont {Zhou}}, \bibinfo {author} {\bibfnamefont {E.}~\bibnamefont {Tsymbal}}, \ and\ \bibinfo {author} {\bibfnamefont {X.}~\bibnamefont {Qiu}},\ }\href {\doibase 10.1038/s41467-022-32179-2} {\bibfield  {journal} {\bibinfo  {journal} {Nat. Commun.}\ }\textbf {\bibinfo {volume} {13}},\ \bibinfo {pages} {4447} (\bibinfo {year} {2022})}\BibitemShut {NoStop}%
\bibitem [{\citenamefont {You}\ \emph {et~al.}(2021)\citenamefont {You}, \citenamefont {Bai}, \citenamefont {Feng}, \citenamefont {Fan}, \citenamefont {Han}, \citenamefont {Zhou}, \citenamefont {Zhou}, \citenamefont {Zhang}, \citenamefont {Chen}, \citenamefont {Pan},\ and\ \citenamefont {Song}}]{You_2021}%
  \BibitemOpen
  \bibfield  {author} {\bibinfo {author} {\bibfnamefont {Y.}~\bibnamefont {You}}, \bibinfo {author} {\bibfnamefont {H.}~\bibnamefont {Bai}}, \bibinfo {author} {\bibfnamefont {X.}~\bibnamefont {Feng}}, \bibinfo {author} {\bibfnamefont {X.}~\bibnamefont {Fan}}, \bibinfo {author} {\bibfnamefont {L.}~\bibnamefont {Han}}, \bibinfo {author} {\bibfnamefont {X.}~\bibnamefont {Zhou}}, \bibinfo {author} {\bibfnamefont {Y.}~\bibnamefont {Zhou}}, \bibinfo {author} {\bibfnamefont {R.}~\bibnamefont {Zhang}}, \bibinfo {author} {\bibfnamefont {T.}~\bibnamefont {Chen}}, \bibinfo {author} {\bibfnamefont {F.}~\bibnamefont {Pan}}, \ and\ \bibinfo {author} {\bibfnamefont {C.}~\bibnamefont {Song}},\ }\href {\doibase 10.1038/s41467-021-26893-6} {\bibfield  {journal} {\bibinfo  {journal} {Nat. Commun.}\ }\textbf {\bibinfo {volume} {12}},\ \bibinfo {pages} {6524} (\bibinfo {year} {2021})}\BibitemShut {NoStop}%
\bibitem [{\citenamefont {Engel}\ \emph {et~al.}(2007)\citenamefont {Engel}, \citenamefont {Rashba},\ and\ \citenamefont {Halperin}}]{Engel_2007}%
  \BibitemOpen
  \bibfield  {author} {\bibinfo {author} {\bibfnamefont {H.-A.}\ \bibnamefont {Engel}}, \bibinfo {author} {\bibfnamefont {E.~I.}\ \bibnamefont {Rashba}}, \ and\ \bibinfo {author} {\bibfnamefont {B.~I.}\ \bibnamefont {Halperin}},\ }\href {\doibase 10.1103/PhysRevLett.98.036602} {\bibfield  {journal} {\bibinfo  {journal} {Phys. Rev. Lett.}\ }\textbf {\bibinfo {volume} {98}},\ \bibinfo {pages} {036602} (\bibinfo {year} {2007})}\BibitemShut {NoStop}%
\bibitem [{\citenamefont {Ma}\ \emph {et~al.}(2021)\citenamefont {Ma}, \citenamefont {Hu}, \citenamefont {Li}, \citenamefont {Liu}, \citenamefont {Yao}, \citenamefont {Jia},\ and\ \citenamefont {Liu}}]{Ma_2021}%
  \BibitemOpen
  \bibfield  {author} {\bibinfo {author} {\bibfnamefont {H.-Y.}\ \bibnamefont {Ma}}, \bibinfo {author} {\bibfnamefont {M.}~\bibnamefont {Hu}}, \bibinfo {author} {\bibfnamefont {N.}~\bibnamefont {Li}}, \bibinfo {author} {\bibfnamefont {J.}~\bibnamefont {Liu}}, \bibinfo {author} {\bibfnamefont {W.}~\bibnamefont {Yao}}, \bibinfo {author} {\bibfnamefont {J.-F.}\ \bibnamefont {Jia}}, \ and\ \bibinfo {author} {\bibfnamefont {J.}~\bibnamefont {Liu}},\ }\href {\doibase 10.1038/s41467-021-23127-7} {\bibfield  {journal} {\bibinfo  {journal} {Nat. Commun.}\ }\textbf {\bibinfo {volume} {12}},\ \bibinfo {pages} {2846} (\bibinfo {year} {2021})}\BibitemShut {NoStop}%
\bibitem [{\citenamefont {Wu}\ \emph {et~al.}(2024)\citenamefont {Wu}, \citenamefont {Deng}, \citenamefont {Yin}, \citenamefont {Tong}, \citenamefont {Tian},\ and\ \citenamefont {Zhang}}]{Wu_2024}%
  \BibitemOpen
  \bibfield  {author} {\bibinfo {author} {\bibfnamefont {Y.}~\bibnamefont {Wu}}, \bibinfo {author} {\bibfnamefont {L.}~\bibnamefont {Deng}}, \bibinfo {author} {\bibfnamefont {X.}~\bibnamefont {Yin}}, \bibinfo {author} {\bibfnamefont {J.}~\bibnamefont {Tong}}, \bibinfo {author} {\bibfnamefont {F.}~\bibnamefont {Tian}}, \ and\ \bibinfo {author} {\bibfnamefont {X.}~\bibnamefont {Zhang}},\ }\href {\doibase 10.1021/acs.nanolett.4c02554} {\bibfield  {journal} {\bibinfo  {journal} {Nano Lett.}\ }\textbf {\bibinfo {volume} {24}},\ \bibinfo {pages} {10534} (\bibinfo {year} {2024})}\BibitemShut {NoStop}%
\bibitem [{\citenamefont {Wang}\ \emph {et~al.}(2023)\citenamefont {Wang}, \citenamefont {Zeng}, \citenamefont {Duan},\ and\ \citenamefont {Huang}}]{Wang_2023}%
  \BibitemOpen
  \bibfield  {author} {\bibinfo {author} {\bibfnamefont {J.}~\bibnamefont {Wang}}, \bibinfo {author} {\bibfnamefont {H.}~\bibnamefont {Zeng}}, \bibinfo {author} {\bibfnamefont {W.}~\bibnamefont {Duan}}, \ and\ \bibinfo {author} {\bibfnamefont {H.}~\bibnamefont {Huang}},\ }\href {\doibase 10.1103/PhysRevLett.131.056401} {\bibfield  {journal} {\bibinfo  {journal} {Phys. Rev. Lett.}\ }\textbf {\bibinfo {volume} {131}},\ \bibinfo {pages} {056401} (\bibinfo {year} {2023})}\BibitemShut {NoStop}%
\end{thebibliography}
%

\end{document}